\documentclass[twocolumn,amsmath,amssymb]{revtex4}
\usepackage[dvips]{graphicx}
\usepackage{dcolumn}
\usepackage{bm}

\begin{document}

\title{Probing high-density symmetry energy using heavy-ion collisions at intermediate energies}
\author{Gao-Chan Yong, Ya-Fei Guo} %\email{yonggaochan@impcas.ac.cn}
\affiliation{$^1$Institute of Modern Physics, Chinese Academy of Sciences, Lanzhou 730000, China\\
$^2$School of Nuclear Science and Technology, University of Chinese Academy of Sciences, Beijing 100049, China}

\begin{abstract}

The nuclear symmetry energy, which describes the energy difference of per proton and neutron in nuclear matter, has been extensively studied within the last two decades. Around saturation density, both the value and the slope of the nuclear symmetry energy have been roughly constrained, its high-density behavior is now still in argument. Probing high-density symmetry energy at terrestrial laboratories is being carried out at facilities that offer radioactive beams worldwide. While relevant experiments are being conducted, we theoretically developed more advanced isospin-dependent transport model including new physics such as nucleon-nucleon short-range correlations and in-medium isospin-dependence of baryon-baryon scattering cross section. New sensitive probes of high-density symmetry energy are provided, such as squeezed-out neutron to proton ratio, photon and light cluster as well as the production of mesons with strangeness or hidden strangeness. The blind spots of probing the high-density symmetry energy by sensitive observable are demonstrated. Model dependence of frequently used sensitive probes of the symmetry energy has been studied thoroughly based on different transport models.
A qualitative observable of neutron to proton ratio at high emitting energy is proposed to probe the high-density symmetry energy qualitatively. The probed density regions of the symmetry energy are carefully studied. Effects of nucleon-nucleon short-range correlations on the some sensitive observables of the symmetry energy in heavy-ion collisions are explored carefully.
Probing the curvature of the symmetry energy by involving the slope information of the symmetry energy at saturation point in the transport model is proposed. Besides constraining the high-density symmetry energy using heavy-ion collisions, a lot of neutron-star related observations from heaven may also be used to constrain the high-density symmetry energy.

\end{abstract}

\maketitle

\section{Introduction}

\begin{figure}[th]
\begin{center}
\includegraphics[width=0.5\textwidth]{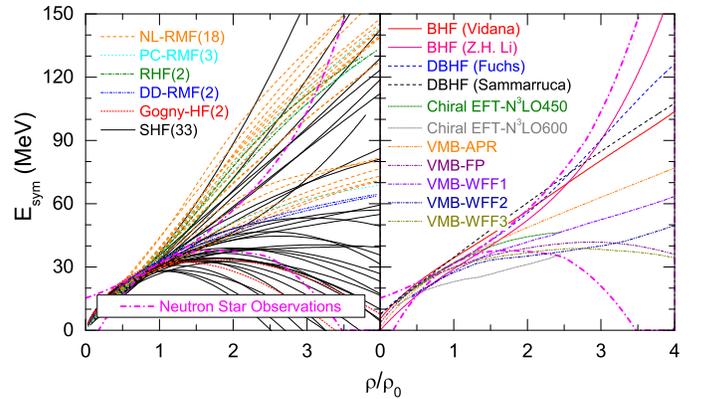}
\end{center}
\caption{Nuclear symmetry energy given by different nuclear many-body approaches in comparison with the constraining boundaries extracted from studying properties of neutron stars. Taken from Ref. \cite{epjareview19}.} \label{unesym}
\end{figure}

To investigate properties of nuclei far from stability and neutron-star-like objects in heaven, the concept \emph{asymmetric nuclear matter} holding unequal numbers of neutron and proton is frequently mentioned.
To describe the properties of asymmetric nuclear matter, the equation of state of nuclear matter is frequently used.
The equation of state (EoS) of
nuclear matter at density $\rho$ and isospin asymmetry
$\delta$ ($\delta=(\rho_n-\rho_p)/(\rho_n+\rho_p)$) usually reads
as \cite{esym91,li08,bar05}
\begin{equation}
E(\rho ,\delta )=E(\rho ,0)+E_{\text{sym}}(\rho )\delta ^{2}+\mathcal{O}%
(\delta ^{4}),
\end{equation}%
where $E_{\text{sym}}(\rho)$ is the nuclear symmetry energy. Apparently, the nuclear symmetry energy describes the change of the single nucleonic energy
of nuclei or nuclear matter when replacing protons with neutrons.
The EoS of isospin symmetric nuclear matter $E(\rho, 0)$ is relatively well
constrained \cite{pawl2002} while the EoS of isospin
asymmetric nuclear matter, especially the high-density symmetry energy, as shown in Fig.\ \ref{unesym}, is still a subject of debate \cite{Guo14,epjareview19,chennpr14,lianpr19,yongnpr09,xucnpr17,zhangyxnpr11}.

The nuclear symmetry energy not only plays crucial roles in nuclear physics
\cite{li08,bar05} but also, in a density range of 0.1 $\sim$ 10 times
nuclear saturation density, affects the
birth of neutron stars and supernova neutrinos \cite{Sumiyoshi95},
the cooling rates and the thickness of the crust of neutron stars, the mass-radius relationship and the moment of inertia of neutron stars \cite{Sum94,Lat04,Ste05a,Lattimer14,wendhnpr15}. The
nuclear symmetry energy also plays important roles in the evolution
of core-collapse supernova \cite{Fischer14} and astrophysical
r-process nucleosynthesis \cite{Nikolov11,rpn1,rpn2,rpn3}, the gravitational-wave frequency \cite{gwf,gwf2} and the gamma-ray bursts \cite{grb} in neutron star mergers \cite{GWth,GW170817}.

Around saturation density both the value of the nuclear symmetry energy (about $31.7\pm3.2$ MeV) and the magnitude of its slope (about $58.7\pm28.1$ MeV) have been roughly determined from a series of concerning analyses \cite{lihan13,Oertel17}. While recent studies on  nuclear experimental measurements from the FOPI and FOPI-LAND by different groups give divergent high-density symmetry energies \cite{isog2015,xiao09,fengplb,russplb,cozma13,xieplb,wangprc}.
The in-medium effects \cite{WMGuo15,hongj2014,xuj2013}, the isospin dependence of strong interactions \cite{yong2011}, the short-range correlations \cite{VRP72,xuc2011} may affect the  interpretation of nuclear experimental data by transport models.

The high-density symmetry energy could be roughly constrained by studying neutron-star merger event GW1760817 since it plays a role in the deformation of dense nuclear matter in neutron-star merger event \cite{epjareview19}. To confine the high-density symmetry energy more directly, currently many terrestrial
experiments are being carried out using a wide variety
of advanced facilities, as done at the GSI Facility for Antiproton and
Ion Research (FAIR) in Germany \cite{fopi16}, the Cooling Storage Ring on the Heavy Ion Research Facility HIRFL-CSR at IMP in China \cite{csr} and the Facility for Rare Isotope Beams (FRIB) in the Untied States \cite{frib}, the Radioactive Isotope Beam Facility (RIBF) at RIKEN in Japan \cite{sep,shan15,ribf} and the Rare Isotope Science Project (RISP) in Korea \cite{korea}.

\section{The isospin-dependent Boltzmann-Uehling-Uhlenbeck (IBUU) transport model}

The Boltzmann-Uehling-Uhlenbeck (BUU) equation denotes time
evolution of single particle phase space distribution function
$f(\vec{r},\vec{p},t)$, which reads \cite{bert88}
\begin{equation}
\frac{\partial f}{\partial
t}+\nabla_{\vec{p}}E\cdot\nabla_{\vec{r}}f-\nabla_{\vec{r}}E\cdot\nabla_{\vec{p}}f=I_{c}.
\label{IBUU}
\end{equation}
The left-hand side of
Eq.~(\ref{IBUU}) describes time evolution of the particle phase
space distribution function due to its transport and mean field,
and the right-hand side accounts for the
modification of the phase space distribution function by elastic and
inelastic two body collisions. $E$ denotes a particle's total energy, which is equal to
kinetic energy $E_{kin}$ plus its average potential energy $U$.
The mean-field potential $U$ of the single particle depends
on its position and momentum of the particle as well as its local asymmetry of medium and is given self-consistently by its phase space distribution function $f(\vec{r},\vec{p},t)$.
The BUU model is one of the most extensively used transport models which describes nucleus-nucleus collisions at different energy regions. At intermediate energies, its main ingredients are single particle potential and particle-particle scattering cross sections as well as pauli-blockings of fermions.

To study the effects of symmetry energy in heavy-ion collisions, isospin dependence is extensively involved into different parts of the BUU transport model.
The isospin- and momentum-dependent single nucleon mean-field
potential reads \cite{li08,das03,yong20171}
\begin{eqnarray}
U(\rho,\delta,\vec{p},\tau)&=&A_u(x)\frac{\rho_{\tau'}}{\rho_0}+A_l(x)\frac{\rho_{\tau}}{\rho_0}\nonumber\\
& &+B\Big(\frac{\rho}{\rho_0}\Big)^{\sigma}(1-x\delta^2)-8x\tau\frac{B}{\sigma+1}\frac{\rho^{\sigma-1}}{\rho_0^\sigma}\delta\rho_{\tau'}\nonumber\\
& &+\frac{2C_{\tau,\tau}}{\rho_0}\int
d^3\,p'\frac{f_\tau(\vec{r},\vec{p^{'}})}{1+(\vec{p}-\vec{p^{'}})^2/\Lambda^2}\nonumber\\
& &+\frac{2C_{\tau,\tau'}}{\rho_0}\int
d^3\,p'\frac{f_{\tau'}(\vec{r},\vec{p^{'}})}{1+(\vec{p}-\vec{p^{'}})^2/\Lambda^2},
\label{buupotential}
\end{eqnarray}
where $\rho_0$ denotes the saturation density, $\tau, \tau'$=1/2(-1/2) is for neutron (proton).
$\rho_n$, $\rho_p$ denote neutron and proton densities,
respectively. The parameter values $A_u(x)$ = 33.037 - 125.34$x$
MeV, $A_l(x)$ = -166.963 + 125.34$x$ MeV, B = 141.96 MeV,
$C_{\tau,\tau}$ = 18.177 MeV, $C_{\tau,\tau'}$ = -178.365 MeV, $\sigma =
1.265$, and $\Lambda = 630.24$ MeV/c.
With these settings, the empirical values of nuclear matter at normal density are reproduced, i.e., the saturation density $\rho_{0}$ = 0.16 fm$^{-3}$, the binding energy $E_{0}$ = -16 MeV, the incompressibility $K_{0}$ = 230 MeV \cite{Oertel17, k230}, the isoscalar effective mass
$m_{s}^{*} = 0.7 m$ \cite{pawel00}, the single-particle potential
$U^{0}_{\infty}$ = 75 MeV at infinitely large nucleon momentum at
saturation density in symmetric nuclear matter, the symmetry
energy $E_{\rm sym}(\rho_0) = 30$ MeV \cite{yong20171}.
\begin{figure}[th]
\centering
\includegraphics[width=0.5\textwidth]{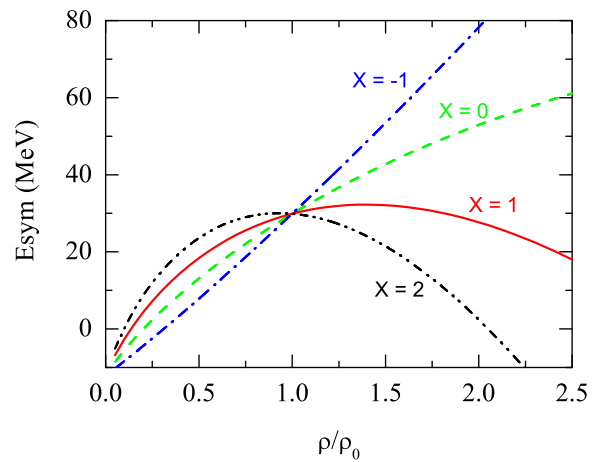}
\caption{Density-dependent symmetry energy with
different $x$ parameters. Taken from Ref. \cite{yong20171}.}\label{esym}
\end{figure}
In Eq.~(\ref{buupotential}), different symmetry energy's stiffness parameters $x$
can be used in different density regions to mimic
different density-dependent symmetry energies. Fig.~\ref{esym} shows the used symmetry energy derived from the single particle potential Eq.~(\ref{buupotential}) with different $x$ parameters at low and high densities.

The isospin-dependent baryon-baryon ($BB$) scattering cross section in medium $\sigma
_{BB}^{medium}$ is reduced compared with their free-space value
$\sigma _{BB}^{free}$ by a factor of \cite{xcross,yong20171}
\begin{eqnarray}
R^{BB}_{medium}(\rho,\delta,\vec{p})&\equiv& \sigma
_{BB_{elastic, inelastic}}^{medium}/\sigma
_{BB_{elastic, inelastic}}^{free}\nonumber\\
&=&(\mu _{BB}^{\ast }/\mu _{BB})^{2},
\end{eqnarray}
where $\mu _{BB}$ and $\mu _{BB}^{\ast }$ are the reduced masses
of the colliding baryon pairs in free space and medium,
respectively. The effective mass of baryon in isospin asymmetric nuclear matter
is expressed as \cite{yong20171}
\begin{equation}
\frac{m_{B}^{\ast }}{m_{B}}=1/\Big(1+\frac{m_{B}}{p}\frac{%
dU}{dp}\Big).
\end{equation}
More details on the present used model can be found in Ref.~\cite{yong20171}.

\section{Sensitive probes of nuclear symmetry energy in heavy-ion collisions}

To constrain the density-dependent symmetry energy by heavy-ion collisions, symmetry energy sensitive probes are usually used as counterparts to compare with relevant experimental data.
Since the symmetry potential has opposite actions for neutrons and protons
and the value of symmetry potential is always quite small compared to
the isoscalar potential, to enlarge the effects of symmetry energy, most observables
in fact use differences or ratios of isospin multiplets of baryons,
mirror nuclei and mesons, such as the neutron/proton ratio, neutron-proton differential flow,
neutron-proton correlation function, $t$/$^{3}$He, $\pi ^{-}/\pi ^{+}$, $\Sigma
^{-}/\Sigma ^{+}$ and $K^{0}/K^{+}$ ratios, etc \cite{li08,bar05}.

\begin{figure}[th]
\begin{center}
\includegraphics[width=0.5\textwidth]{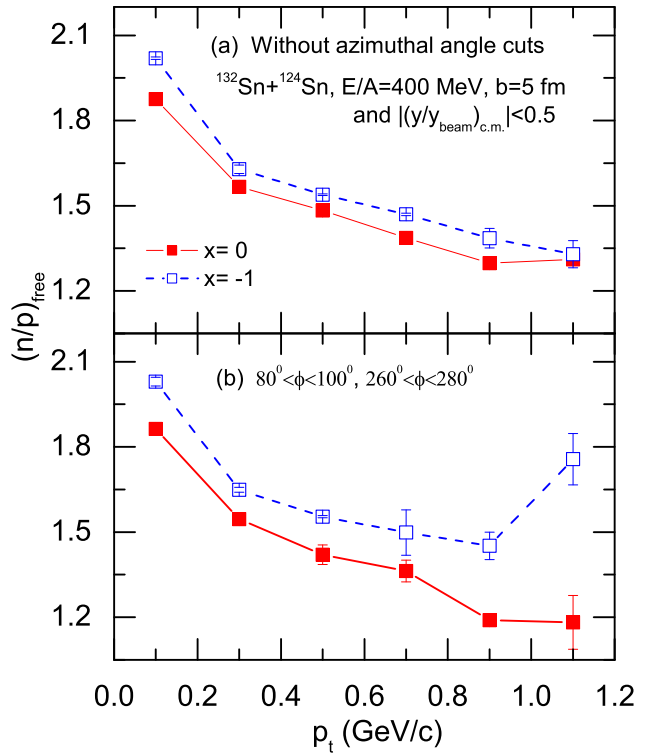}
\end{center}
\caption{Effects of symmetry energy on the non-squeezed-out (panel (a)) and squeezed-out (panel (b)) neutron to proton ratios in $^{132}$Sn+$^{124}$Sn reaction at $400$
MeV/nucleon. Taken from Ref. \cite{yong07}.} \label{rnpa}
\end{figure}
Squeezed-out nucleons emitted in the direction of perpendicular to the reaction plane
in semi-central heavy-ion collisions carry more information on the property of dense matter.
Fig.\ \ref{rnpa} demonstrates transverse momentum dependent
neutron/proton ratio of mid-rapidity nucleons emitted in the
direction perpendicular to the reaction plane \cite{yong07}. It
is seen in panel (b) that the symmetry energy
effect on the n/p ratio increases with the increasing transverse
momentum $p_t$ and the effect can
be as high as 40\% at high transverse momenta. The high $p_t$ nucleons most likely originate from the high density region in the early stage in heavy-ion
collisions thus is more sensitive to the high-density
symmetry energy. The n/p ratio of free nucleons at mid-rapidity
without azimuthal angle cut (shown in panel (a)) is much less sensitive to the
symmetry energy \cite{yong07}.

\begin{figure}[th]
\begin{center}
\includegraphics[width=0.4\textwidth]{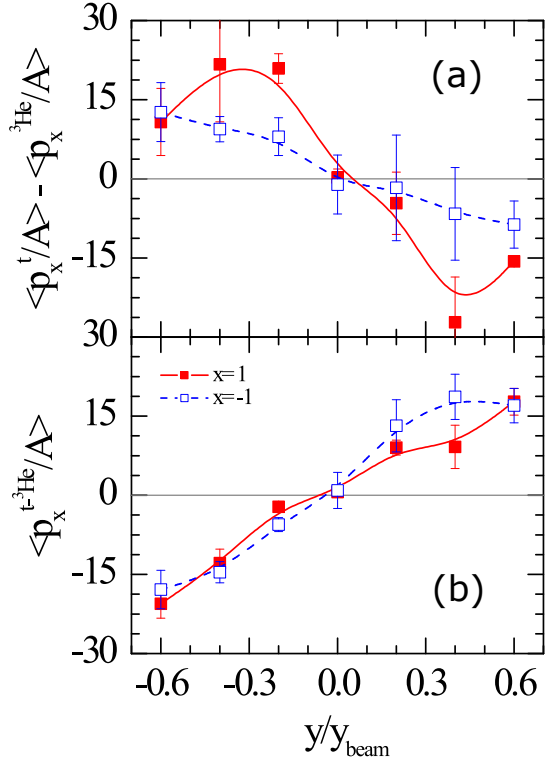}
\end{center}
\caption{Effects of symmetry energy on the triton-$^{3}$He relative (panel (a)) and differential (panel (b)) flows as a function of rapidity in the semi-central reaction of $^{132}Sn+^{124}Sn$
at 400 MeV/nucleon. Taken from Ref. \cite{cluster09}.} \label{clusterflow}
\end{figure}
It is practically difficult to measure observables involving neutrons. One
question frequently being asked is whether the
triton-$^{3}$He (t -$^{3}$He) pair may carry the information of nuclear symmetry energy.
It is normally difficult to extract reliable information about the symmetry
energy from the individual flows of triton and $^{3}$He clusters.
To decrease effects of the isoscalar potential
while enhancing effects of the isovector potential are
helpful. Fig.\ \ref{clusterflow} shows the triton-$^{3}$He relative and
differential flows as a function of rapidity \cite{cluster09}. From panel (a) of Fig.\
\ref{clusterflow}, it is shown that the triton-$^{3}$He relative
flow is very sensitive to the symmetry energy. Effects of the
symmetry energy on the differential flow are shown in panel (b),
is relatively small. Therefore triton-$^{3}$He relative flow is a potential probe used
to detect the symmetry energy in heavy-ion collisions.

\begin{figure}
\begin{center}
\includegraphics[width=0.45\textwidth]{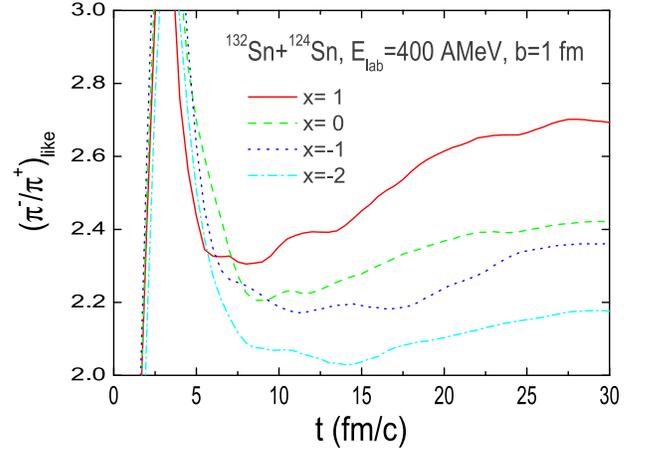}
\end{center}
\caption{Effects of symmetry energy on the $(\pi^-/\pi^+)_{like}$ ratio in the central $^{132}Sn+^{124}Sn$ reaction
at 400 MeV/nucleon. Taken from Ref. \cite{lyz05}.} \label{rpitim}
\end{figure}
Charged $\pi^-/\pi^+$ ratio is another sensitive probe of nuclear symmetry energy.
Shown in Fig.\ \ref{rpitim} is $(\pi^-/\pi^+)_{like}$ ratio (=$ \frac{\pi^-+\Delta^-+\frac{1}{3}\Delta^0}
{\pi^++\Delta^{++}+\frac{1}{3}\Delta^+}$) as a function of time \cite{lyz05}. Due to more neutron-neutron scatterings than that of proton-proton when two neutron skins start overlapping at the beginning of the reaction, the value of the $(\pi^-/\pi^+)_{like}$ ratio soars in the early stage of the reaction. The $(\pi^-/\pi^+)_{like}$ ratio saturates at about 25 fm/c indicating that a chemical freeze-out stage has been reached. The sensitivity to the symmetry energy is clearly shown in the final $\pi^-/\pi^+$ ratio.

\begin{figure}[th]
\begin{center}
\includegraphics[width=0.45\textwidth]{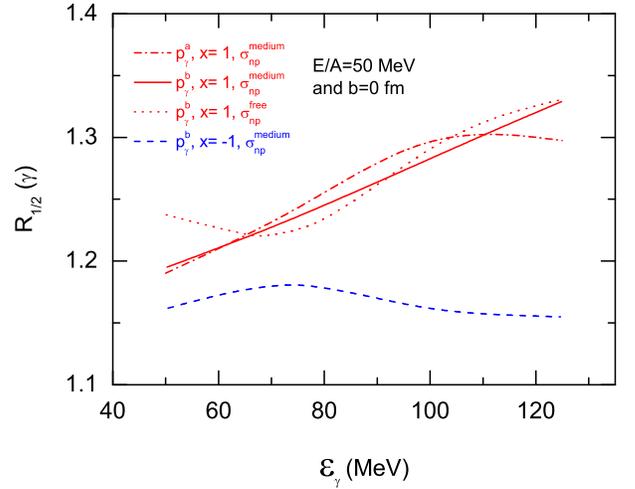}
\end{center}
\caption{Effects of symmetry energy on the spectra ratio of hard photons in
reactions of $^{132}Sn+^{124}Sn$ and $^{112}Sn+^{112}Sn$
at 50 MeV/A, see text for details. Taken from Ref. \cite{photon}.} \label{relat}
\end{figure}
The hadronic probes usually suffer from distortions due to the strong
interactions in the final state. It is preferable to get more clean ways to probe the symmetry energy especially at supranormal densities.
Fig.\ \ref{relat} shows the spectra ratio $R_{1/2}(\gamma)$ of hard photons with
$p^a_{\gamma}(=1.55\times10^{-3}\times\frac{1}{\varepsilon_{\gamma}}(\beta_{i}^{2}+\beta_{f}^{2}))$ and $p^b_{\gamma}$($=2.1\times10^{-6}\frac{(1-y^{2})^{\alpha}}{y}$, $y=\varepsilon_{\gamma}/E_{max}$,
$\alpha=0.7319-0.5898\beta_i$, $\beta_i$ and
$\beta_f$ are the initial and final velocities, $E_{max}$ is the energy
available in the center of mass) production forms \cite{photon}. It is first seen
that calculations with $p^a_{\gamma}$ and
$p^b_{\gamma}$ and in-medium NN cross sections all lead to
about the same $R_{1/2}(\gamma)$ within statistical errors as
expected. The effect of the in-medium NN
cross sections is canceled out in the spectra ratio. Free from the uncertainties of
elementary photon production and the NN cross sections, the observable
$R_{1/2}(\gamma)$ could be considered as a robust probe of the symmetry energy.
Comparing the calculations with soft and stiff symmetries
both using the $p^b_{\gamma}$, it is seen that the
$R_{1/2}(\gamma)$ is very sensitive to the symmetry energy
especially for energetic photons.

\begin{figure}[th]
\begin{center}
\includegraphics[width=0.5\textwidth]{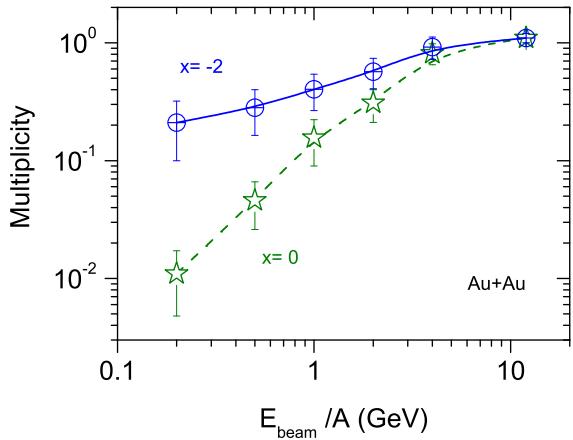}
\end{center}
\caption{Effects of symmetry energy on the multiplicity of $\eta$ production
as a function of incident beam energy in Au+Au reactions. Taken from Ref. \cite{eta13}.} \label{exc}
\end{figure}
In the light of sensitivity of the $\pi^-/\pi^+$ ratio to the symmetry energy at subthreshold energies, the multiplicity of $\eta$ production as a function of incident beam energy is also examined especially below its production threshold.
Compared to pions, $\eta$ mesons experience
weaker final state interactions due to hidden strangeness.
Fig.~\ref{exc} shows $\eta$ multiplicity as a
function of beam energy in Au+Au reaction with soft
($x=0$) and stiff ($x=-2$) symmetry energies \cite{eta13}. One can see that the multiplicity of $\eta$ declines with decreasing beam energy, especially for the
soft symmetry energy. And it saturates at
incident energy of about 10 GeV/nucleon. The effect of nuclear symmetry energy on the $\eta$ production is much more evident than the $\pi^-/\pi^+$ ratio especially in the deeper sub-threshold region.

\section{Blind spots of probing the high-density symmetry energy in heavy-ion collisions}

Based on the chemical equilibrium condition of nuclear matter \cite{chemical2002}, if the symmetry energy does not change with density, the liquid-gas phase transition cannot occur. Therefore, in case the symmetry energy is less density-dependent, the effects of symmetry energy on the final-state observable may disappear.

To show the blind spots of probing the high-density symmetry energy, the neutron to proton ratio
ratio in the central Au+Au reaction at 300 MeV/nucleon is used as an example \cite{yongblind20182}. In the used transport model, the effects of neutron-proton short-range-correlations are taken into account \cite{yong20171,yong20172}. Also the transition momentum of the proton is set to be the same as that of neutron \cite{yong2018}. The in-medium inelastic baryon-baryon collisions and the pion in-medium transport are included as well \cite{yongm2016,yongp2015}. For initialization, nucleon density distribution in colliding nuclei is calculated using the Skyrme-Hartree-Fock with Skyrme M$^{\ast}$ force parameters \cite{skyrme86}. And the proton and neutron momentum distributions with high-momentum tails in initial nucleus are reproduced  \cite{yong20171,yong20172,sci08,sci14,yongcut2017}.

\begin{figure}[th]
\centering
\includegraphics[width=0.55\textwidth]{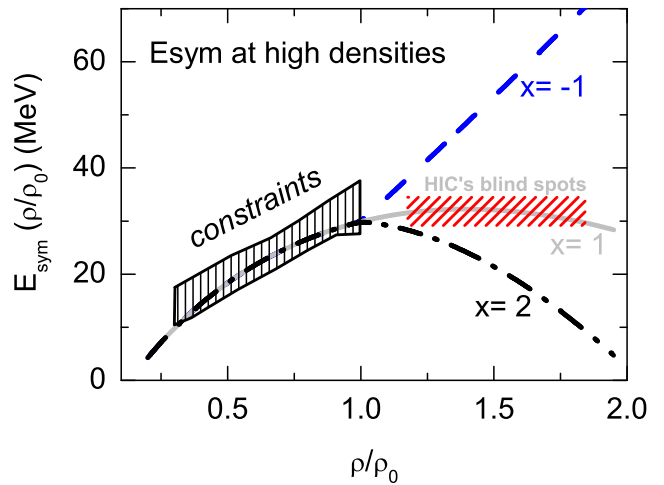}
\caption{The density-dependent symmetry energy at low and high densities. The symmetry energy below saturation density is fixed as experimental constraints while the high-density symmetry energy is alterable. Taken from Ref. \cite{yongblind20182}.} \label{esymbl}
\end{figure}
Since below saturation density, the symmetry energy is roughly constrained \cite{horowitz2014,lwchen2017}, we simulate the low-density symmetry energy with parameter $x$ = 1. From Fig.~\ref{esymbl}, it is seen that
with parameter $x$ = 1 the simulated symmetry energy is well consistent with the current constraints. In the study, we vary the high-density symmetry energy with parameter $x$ in the range of $x$ = -1, 1, 2, which cover the current uncertainties of the high-density symmetry energy \cite{Guo14,lwchen2017}.

\begin{figure}[t]
\centering
\includegraphics[width=0.55\textwidth]{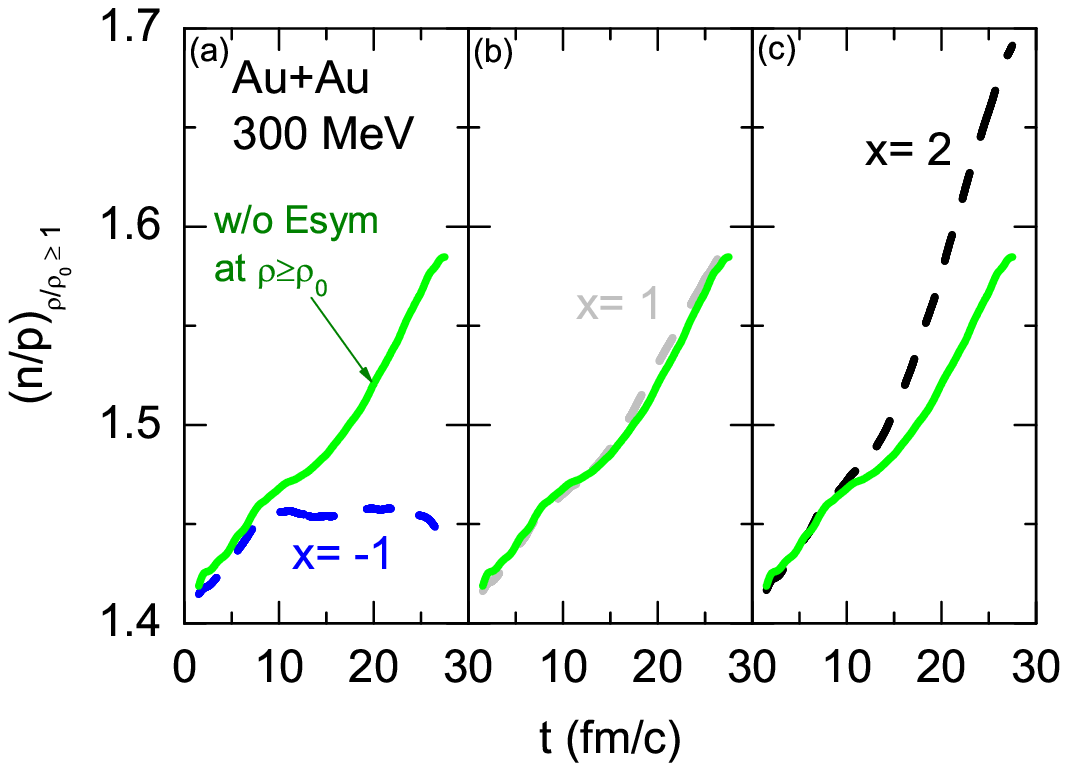}
\caption{Effects of the high-density symmetry energy on the neutron to proton ratio n/p in dense matter ($\rho/\rho_{0}\geq1$) as a function of time in central Au+Au
reaction at 300 MeV/nucleon . The solid line denotes the case without high-density symmetry energy, panels (a), (b), (c) indicate cases with stiff, soft and super-soft symmetry energies, respectively. Taken from Ref. \cite{yongblind20182}.} \label{densernp}
\end{figure}
Fig.~\ref{densernp} shows the effects of high-density symmetry energy on the neutron to proton ratio n/p of dense matter formed in central Au+Au reaction at 300 MeV/nucleon. In the study, the same low-density symmetry energy as shown in Fig.~\ref{esymbl} is used. From Fig.~\ref{densernp} (a) and (c), one sees that the stiffer/soft high-density symmetry energy causes a smaller/large asymmetry of dense matter. While from Fig.~\ref{densernp} (b), it is seen that the less density-dependent high-density symmetry energy ($x$= 1, shown in Fig.~\ref{esymbl}) almost does not affect the asymmetry of dense matter. For isospin-fractionation, there is a chemical equilibrium condition \cite{chemical2002,muller,liko,baran,shi}
\begin{equation}\label{chem}
E_{sym}(\rho_1)\delta_1=E_{sym}(\rho_2)\delta_2,
\end{equation}
where $E_{sym}(\rho_1), E_{sym}(\rho_2)$ denote symmetry energies at different density regions and $\delta_1, \delta_2$ are, respectively, their asymmetries.
In nuclear matter there is a dynamical process of nucleon movement
determined by Eq. (\ref{chem}) according to the
density dependent symmetry energy.
Since the high-density symmetry energy with parameter $x$= 1, $ E_{sym}(\rho_1) \approx E_{sym}(\rho_2)$, the asymmetry of dense matter $\delta_1\approx\delta_2$.
One thus sees in Fig.~\ref{densernp} (b), the less density-dependent high-density symmetry energy almost does not affect the asymmetry of dense matter formed in heavy-ion collisions.

\begin{figure}[t]
\centering
\includegraphics[width=0.55\textwidth]{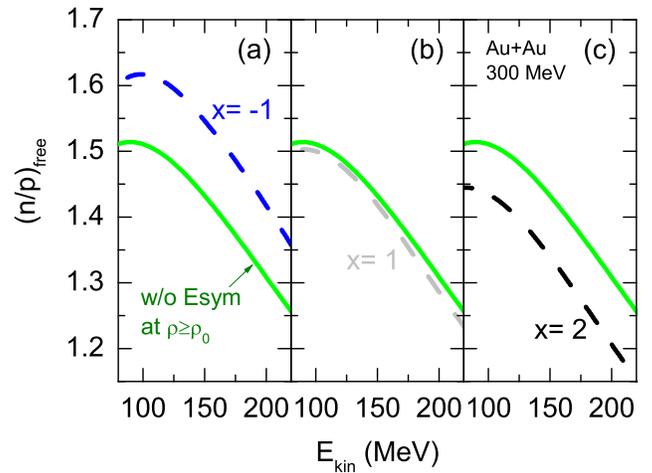}
\caption{Effects of high-density symmetry energy on the free neutron to proton ratio n/p as a function of kinetic energy in central Au+Au
reaction at 300 MeV/nucleon. The solid line denotes the case without high-density symmetry energy, panels (a), (b), (c) indicate cases with stiff, soft and super-soft symmetry energies, respectively. Taken from Ref. \cite{yongblind20182}.} \label{freernp}
\end{figure}
Since the symmetry potential acts directly on nucleons, the neutron/proton ratio n/p of nucleon emissions may be one of the best observables to probe the symmetry energy \cite{BCY2005,li1997}.
Fig.~\ref{freernp} shows the effects of high-density symmetry energy on the kinetic energy distribution of the free neutron to proton ratio n/p in central Au+Au reaction at 300 MeV/nucleon. From Fig.~\ref{freernp} (a), (c), it is seen that the stiffer/soft high-density symmetry energy causes a larger/small n/p ratio. While from Fig.~\ref{freernp} (b)
one sees that the less density-dependent high-density symmetry energy almost does not affect the free n/p ratio. Therefore, the less density-dependent high-density symmetry energy cannot be probed effectively in heavy-ion collisions.

\section{Model dependence of symmetry-energy-sensitive probes and qualitative probe}

There are many factors affecting nuclear reaction
transport simulation, e.g., the initialization, the nucleon-nucleon interaction or single nucleon potential, the nucleon-nucleon scattering cross sections, and the framework of transport models. It is thus necessary to make a study among different models, to see how large the
differences are on the values of isospin sensitive observables.

Within the frameworks of isospin-dependent transport models Boltzmann-Uehling-Uhlenbeck (IBUU04) \cite{lyz05}
and Ultrarelativistic Quantum Molecular Dynamics (UrQMD) \cite{urqmd1,urqmd2}, the model dependences of frequently used isospin-sensitive observables $\pi^{-}/\pi^{+}$ ratio and $n/p$
ratio of free nucleons have been demonstrated \cite{gwm2013}.
\begin{figure}[htb]
\begin{center}
\includegraphics[width=0.5\textwidth]{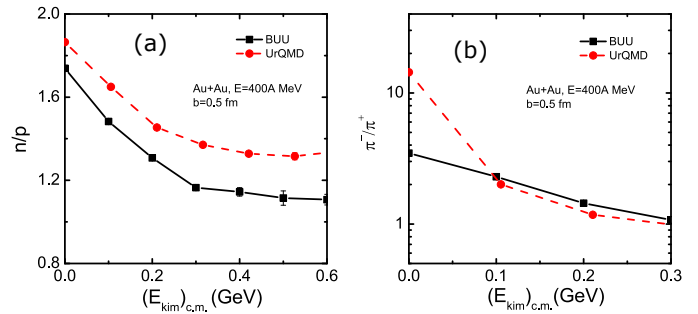}
\end{center}
\caption{Model dependence of $n/p$ ratio of free nucleons and $\pi^-/\pi^+$ ratio as a
function of kinetic energy in $^{197}Au+^{197}Au$ at 400 MeV/nucleon. Panels (a) and (b) show n/p and $\pi^-/\pi^+$ ratio cases, respectively. Taken from Ref.  \cite{gwm2013}.} \label{fig:Fig2}
\end{figure}
Fig.~\ref{fig:Fig2} shows the $n/p$ ratio of free nucleons
and $\pi^-/\pi^+$ ratio as a function of kinetic energy in the
$^{197}Au+^{197}Au$ reaction at a beam energy of 400
MeV/nucleon simulated by the IBUU and the UrQMD models. From panel (a) one can see that both models give the same trend of n/p ratio as a function of nucleonic kinetic energy.
The result of the UrQMD model is overall larger than that
of the IBUU model due to their different forms of symmetry
potential \cite{gwm2013}. From panel (b) of
Fig.~\ref{fig:Fig2}, one can see that there is a cross between the
$\pi^-/\pi^+$ ratios from the UrQMD model and that from the IBUU
model. At lower kinetic energies, the value of $\pi^-/\pi^+$
ratio from the UrQMD is much larger than that from the IBUU model.
\begin{figure}[htb]
\begin{center}
\includegraphics[width=0.5\textwidth]{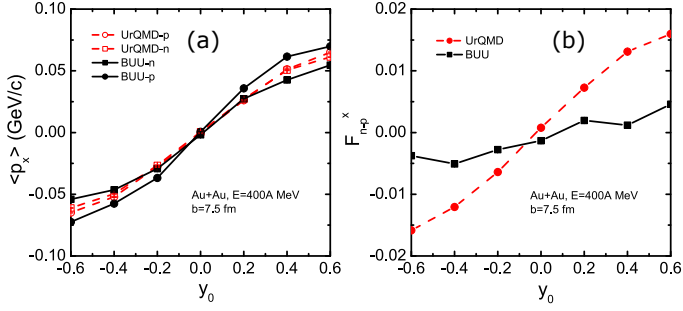}
\end{center}
\caption{Same as Fig.~\ref{fig:Fig2}, but for the rapidity distributions of the transverse flow $<p_x(y)>$ for neutrons and protons (panel (a)) and the neutron-proton differential
transverse flow $F^{x}_{n-p}$ (panel (b)). Taken from Ref. \cite{gwm2013}.} \label{fig:Fig3}
\end{figure}
And from panel (a) of Fig.~\ref{fig:Fig3}, one can see that nucleonic transverse flow given by the IBUU model shows large isospin effect than that with the UrQMD
model. From panel (b) of Fig.~\ref{fig:Fig3}, it is seen that the slope of neutron-proton differential flow is evidently larger for the UrQMD model than that for the IBUU model.

Since most observables of the symmetry energy are model dependent, on the first step, what is crucially needed currently is a qualitative observable to probe whether the symmetry energy at high densities is stiff or soft.

\begin{figure}[htb]
\begin{center}
\includegraphics[width=0.5\textwidth]{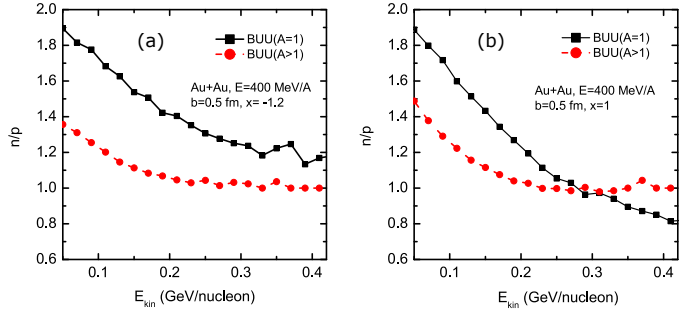}
\end{center}
\caption{kinetic energy distribution of neutron to proton ratio $n/p$ of free (panel (a)) and bound nucleons (panel (b))
in the central $^{197}Au+^{197}Au$ reaction at 400 MeV/A with
positive (left, x= -1.2) and negative (right, x= 1) symmetry potentials at
supra-saturation densities. Given by the IBUU transport
model. Taken from Ref. \cite{Guo14}.} \label{buukinetic}
\end{figure}
\begin{figure}[htb]
\begin{center}
\includegraphics[width=0.5\textwidth]{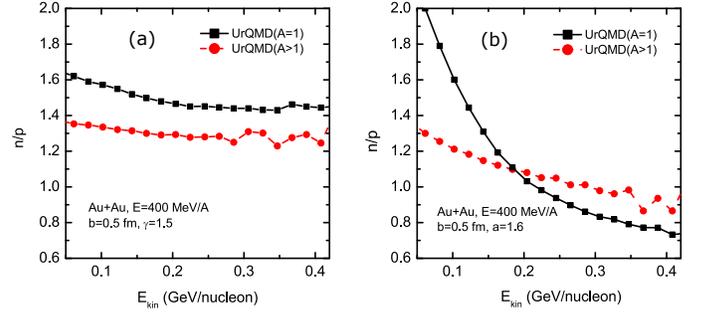}
\end{center}
\caption{Same as Figure~\ref{buukinetic} but given by the UrQMD transport
model with positive (left, $\gamma$= 1.5) and negative (right, a= 1.6) symmetry potentials at
supra-saturation densities. Taken from Ref. \cite{Guo14}.} \label{qmdkinetic}
\end{figure}
Figure~\ref{buukinetic} shows nucleon kinetic energy distribution of
n/p ratios of free (gas) and bound (liquid) nucleons in the
central $^{197}Au+^{197}Au$ reaction at 400 MeV/A
with positive and negative symmetry
potentials at supra-saturation densities given by the IBUU
transport model \cite{Guo14}. From panel (a), it is seen that the value of
$n/p$ of nuclear gas phase is larger than that of liquid
phase in the whole kinetic energy distribution
with the positive symmetry potential. However, it is
interesting to see that the value of $n/p$ of gas phase
is smaller than that of liquid phase at higher kinetic
energies with the negative symmetry potential at
high densities. The same situation is demonstrated in Fig.~\ref{qmdkinetic} with the UrQMD model. The only discrepancy is that the kinetic energy of
transition point given by the UrQMD model is lower than that given by the
IBUU model. With positive/negative symmetry potential at high densities, for energetic nucleons, the value of neutron to proton ratio of free nucleons is larger/smaller than that of
bound nucleon fragments. Compared with extensively studied
quantitative observables of the nuclear symmetry energy, the normal or
abnormal isospin-fractionation of energetic nucleons can be a
qualitative probe of nuclear symmetry energy at high densities.

\section{Determination of the density region of the symmetry energy probed by the $\pi^-/\pi^+$ ratio and nucleon observables}

To investigate the symmetry energy at high densities, the $\pi^-/\pi^+$ ratio has been frequently proposed in the literature \cite{NPDF2,lyz05,y2006}. Aiming at probing the high-density symmetry energy by the $\pi^-/\pi^+$ ratio, Sn+Sn reactions at about 270 MeV/nucleon experiments are being carried out at RIKEN in Japan \cite{shan15,exp2,exp3}. Does the $\pi^-/\pi^+$ ratio in heavy-ion collisions at intermediate energies always probe the symmetry energy above saturation density? Alternatively, in what conditions does the $\pi^-/\pi^+$ ratio probe the symmetry energy at high densities?

\begin{figure}[t!]
\centering
\includegraphics [width=0.5\textwidth]{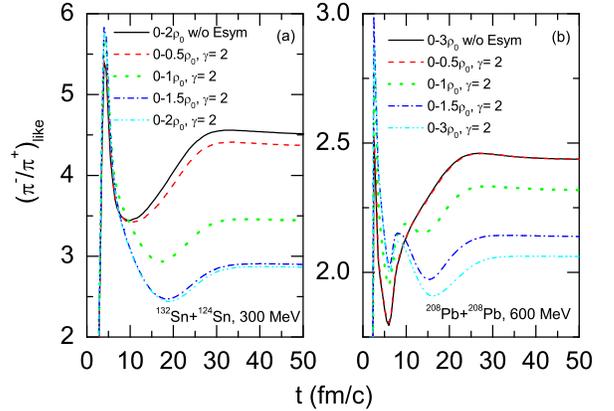}
\caption{\label{rpitime}Evolution of the relative sensitivity of the
symmetry energy observable $(\pi^-/\pi^+)_{like}$ ratio in $^{132}$Sn+$^{124}$Sn at 300 MeV/nucleon (panel (a)) and $^{208}$Pb+$^{208}$Pb at 600 MeV/nucleon (panel (b)). Taken from Ref. \cite{y2019}.}
\end{figure}
To answer the above questions, we studied the decomposition of the sensitivity of the symmetry energy observables. From Figure~\ref{rpitime}(a), it is demonstrated that, in the medium-mass nuclei $^{132}$Sn+$^{124}$Sn central reaction at 300 MeV/nucleon, the effects of the symmetry energy in the densities below 0.5$\rho_{0}$ are larger than that in the densities above 1.5$\rho_{0}$. Below saturation density the effects of the symmetry energy are roughly equal to that above saturation density. However, from Figure~\ref{rpitime}(b), one can see that, the effects of the symmetry energy in the densities below 0.5$\rho_{0}$ are obviously smaller than that in the densities above 1.5$\rho_{0}$. The effects of the symmetry energy below saturation density are obviously smaller than that above saturation density.
Therefore, to probe the high-density symmetry energy by the $\pi^-/\pi^+$ ratio, heavy system and at relatively higher incident beam energies is a preferable.

Similar studies were also carried out for some other frequently discussed observables \cite{fan2018}. The symmetry energy sensitive observable n/p ratio in the $^{132}$Sn+$^{124}$Sn reaction at 0.3 GeV/nucleon is shown to be sensitive to the symmetry energy below $1.5\rho_0$. Nucleon elliptic flow can probe the symmetry energy from low to high densities when changing the beam energies from 0.3 to 0.6 GeV/nucleon in the semi-central $^{132}$Sn+$^{124}$Sn reaction. And nucleon transverse and elliptic flows in the semi-central $^{197}$Au+$^{197}$Au reaction at 0.6 GeV/nucleon are more sensitive to the high-density symmetry energy.

\section{Effects of short-range correlations in transport model}

\begin{figure}[th]
\centering
\includegraphics[width=0.5\textwidth]{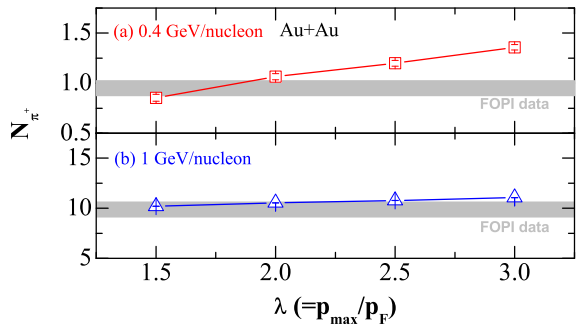}
\caption{Effects of short-range correlations on the number of produced $\pi^{+}$ meson production in Au + Au collisions at 0.4 (panel (a)) and 1 GeV/nucleon (panel (b)) beam energies. Taken from Ref. \cite{yongcut2017}.}
\label{pionx}
\end{figure}
Recent proton-removal experiments showed that only about 80\%
nucleons participate in the independent particle motion
\cite{e93,e96,sci08}. Also it is shown that nucleons in nuclei can form pairs with larger
relative momenta and smaller center-of-mass momenta
\cite{pia06,sh07}. This is interpreted by
the nucleon-nucleon short-range tensor interaction \cite{tenf05,tenf07}.
The nucleon-nucleon short-range correlations (SRC) in nuclei cause a high-momentum
tail (HMT) in single-nucleon momentum distribution above Fermi momentum
\cite{bethe71,anto88,Rios09,yin13,Claudio15}.
And the high-momentum tail's shape is almost identical for all nuclei
\cite{Ciofi96,Fantoni84,Pieper92,egiyan03}, i.e., roughly exhibits
a $C/k^{4}$ distribution \cite{hen14,sci14,henprc15,liba15}.

The short-range correlations in nuclei surely affect $\pi$, $K$, $\eta$ and nucleon emission in heavy-ion collisions. Fig.~\ref{pionx} shows $\pi^{+}$ production as a
function of high-momentum tail cutoff parameter $\lambda$ in Au + Au collisions at 0.4 and 1 GeV/nucleon incident beam energies \cite{yongcut2017}. Larger high-momentum cutoff parameter causes larger nucleon average kinetic energy, especially
proton average kinetic energy, thus more $\pi^{+}$'s are produced. As incident beam energy increases, the initial movement of nucleons in nuclei becomes less important in nucleus-nucleus
collisions.

\begin{figure}[th]
\begin{center}
\includegraphics[width=0.5\textwidth]{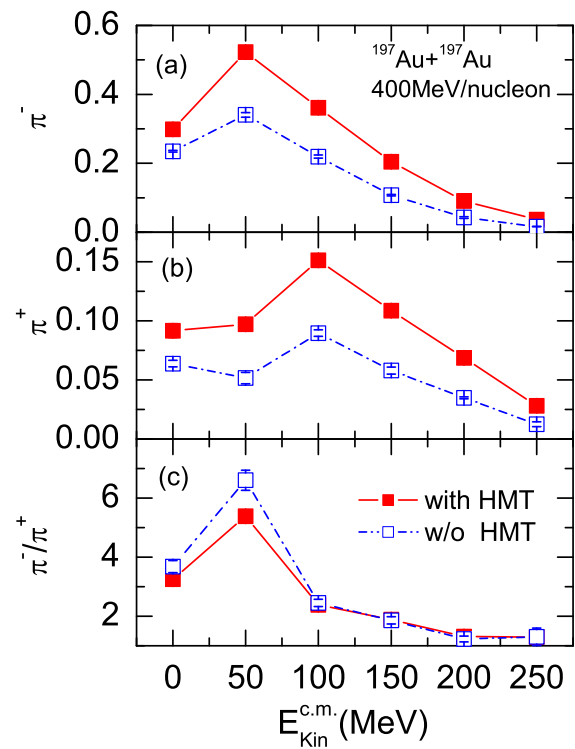}
\end{center}
\caption{Effects of short-range correlations on the kinetic energy distributions of $\pi^{-}$ (panel (a)), $\pi^{+}$ (panel (b)) and $\pi^{-}/\pi^{+}$ ratio (panel (c)) in the reaction of $^{197}\rm {Au}+^{197}\rm {Au}$ at 400 MeV/nucleon. Taken from Ref. \cite{zhangf}.}
\label{Rpion}
\end{figure}
\begin{figure}[th]
\begin{center}
\includegraphics[width=0.5\textwidth]{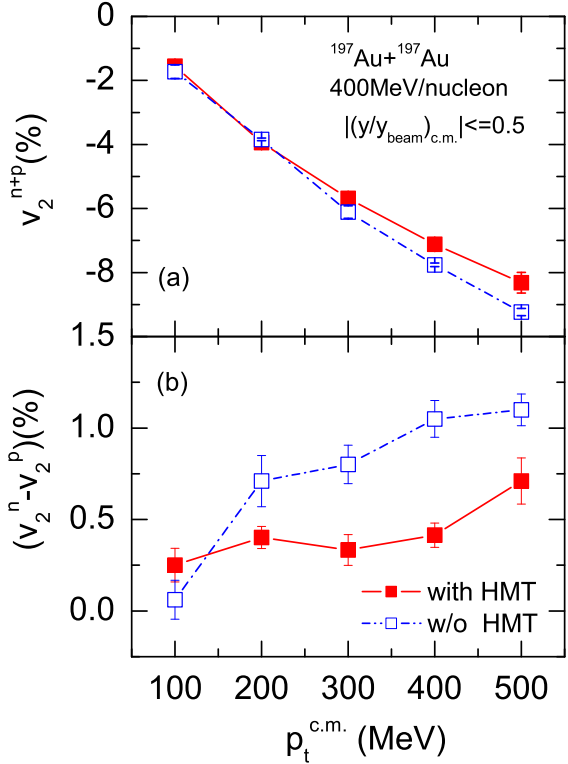}
\end{center}
\caption{Effects of short-range correlations on the nucleon elliptic flow (panel (a)) and the
difference of neutron and proton elliptic flow (panel (b)) in the semi-central reaction
of $^{197}\rm {Au}+^{197}\rm {Au}$ at 400 MeV/nucleon. Taken from Ref. \cite{zhangf}.}
\label{v2ptnp}
\end{figure}
Fig.\ \ref{Rpion} shows effects of short-range correlations on the kinetic energy distributions of $\pi^{-}$, $\pi^{+}$ as well as $\pi^{-}/\pi^{+}$ ratio in the semi-central reaction
$^{197}\rm {Au}+^{197}\rm {Au}$ at 400 MeV/nucleon \cite{zhangf}. It is clearly demonstrated that the kinetic energy distributions of both $\pi^{-}$ and $\pi^{+}$ are sensitive to the HMT. The ratio of $\pi^{-}/\pi^{+}$ is sensitive to the HMT except in the high kinetic energy region. The short-range corrections increase kinetic energies of neutrons and protons, thus more pion's are produced. With the short-range correlations, protons have a larger probability than neutrons to have larger momenta, one thus sees a lower value of the $\pi^{-}/\pi^{+}$ ratio with the HMT.
From Fig.\ \ref{v2ptnp}, one can see that the effects of the HMT on the difference of
neutron and proton elliptic flows ($v_{2}^{n}-v_{2}^{p}$) is also crucial \cite{zhangf}.

In neutron-rich matter, neutron and proton may have very different Fermi momenta. If each correlated neutron and proton have $1/k^{4}$ distributions starting from their respective Fermi momenta, based on the n-p dominance model \cite{sci14}, the correlated neutron and proton cannot have the same momentum magnitude. It is thus reasonable to think that in neutron-rich matter proton has the same transition momentum as that of neutron.
\begin{figure}[th]
\centering
\includegraphics[width=0.5\textwidth]{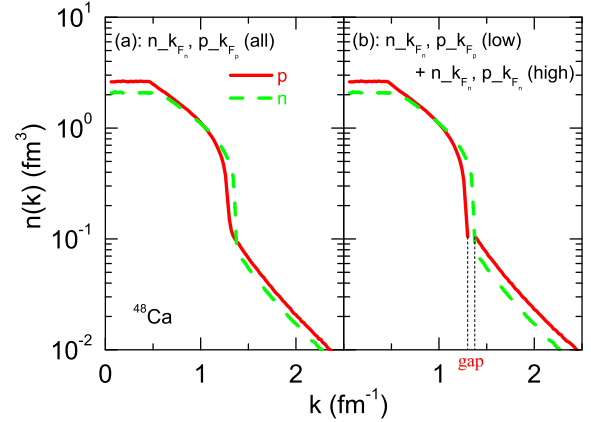}
\caption{ Momentum distribution of nucleon with different starting points of protons in $^{48}$Ca. Panels (a), (b) show cases without and with proton momentum gap in nucleon momentum distributions, respectively. Taken from Ref. \cite{jump18}.} \label{npdis}
\end{figure}
\begin{figure}[th]
\centering
\includegraphics[width=0.5\textwidth]{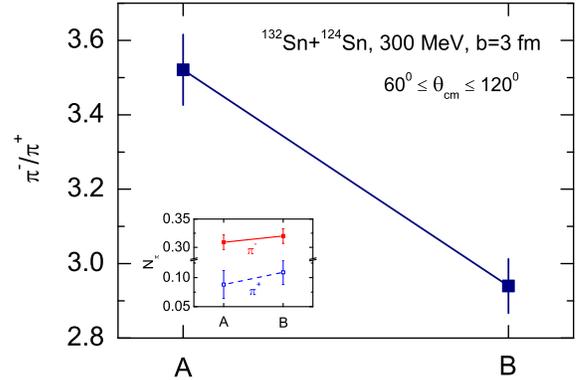}
\caption{ The ratio of $\pi^{-}/\pi^{+}$ in $^{132}$Sn+$^{124}$Sn reactions at 300 MeV/nucleon with different proton starting momenta in the HMT. Taken from Ref. \cite{jump18}.} \label{pion}
\end{figure}
Fig.~\ref{npdis} shows nucleon momentum distribution in $^{48}$Ca. One can see that a HMT above the nuclear Fermi momentum is produced. Proton has greater probability than neutron to have momenta greater than the nuclear Fermi momentum. Compared case A with case B, it is seen that with the starting point of neutron Fermi momentum, proton has even more greater probability to have high momenta. The dynamics of heavy-ion collisions could be affected by proton momentum gap. Fig.~\ref{pion} demonstrates the ratio of $\pi^{-}/\pi^{+}$ in $^{132}$Sn+$^{124}$Sn reactions with different proton starting momenta in the HMT. As expected, there is an evident decline of the value of $\pi^{-}/\pi^{+}$ ratio when varying the starting momentum of proton in the HMT. From the inserted figure, it is seen that mainly the $\pi^{+}$ production is affected. To interpret related experiments at Radioactive Isotope Beam Facility (RIBF) at RIKEN in Japan \cite{shan15,exp2}, it is necessary to study how the $\pi^{-}/\pi^{+}$ ratio is affected by the proton momentum gap.

\section{Cross-checking the symmetry energy at high densities}

\begin{figure}[tbh]
\centering
\includegraphics[width=0.5\textwidth]{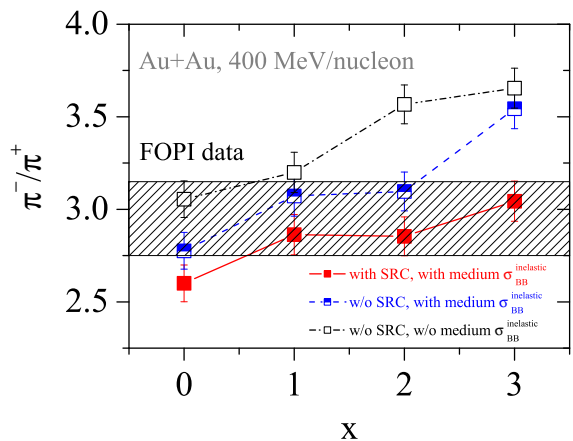}
\caption{Effects of symmetry energy on the $\pi^{-}/\pi^{+}$ ratio in Au+Au reaction at 400 MeV/nucleon with different short-range correlations and in-medium cross section. Taken from Ref. \cite{yongm2016}.} \label{rpion}
\end{figure}
\begin{figure}[tbh]
\centering
\includegraphics[width=0.5\textwidth]{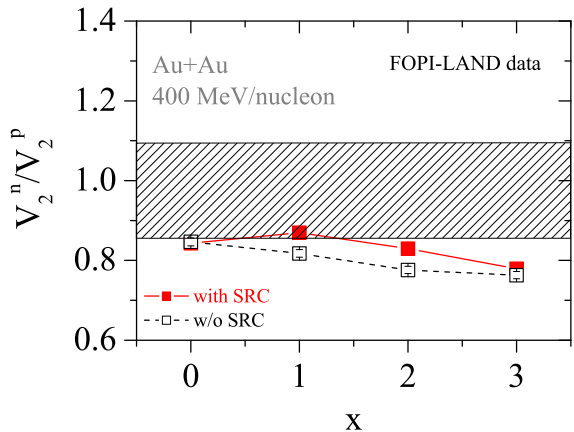}
\caption{Effects of symmetry energy on the ratio of neutron and proton elliptic flows $V_{2}^{n}/V_{2}^{p}$ in Au+Au reaction at 400 MeV/nucleon. Taken from Ref. \cite{yongm2016}.}
\label{v2rnp}
\end{figure}
Since the constraints on the symmetry energy are always model dependent, it is instructive to
make cross-check of the symmetry energy using different probes.
Fig.~\ref{rpion} shows the $\pi^{-}/\pi^{+}$ ratio given by the IBUU model. It is seen
that larger $\pi^{-}/\pi^{+}$ ratio corresponds
softer symmetry energy. One can also see that both the short-range correlations and in-medium cross section affect the value of $\pi^{-}/\pi^{+}$ ratio evidently. From Fig.~\ref{rpion}, it is concluded that the FOPI pion data supports a softer symmetry energy ($x$ = 1, 2, even $x$ = 3).

Fig.~\ref{v2rnp} shows comparison of simulated nucleon elliptic flow
and experimental data. Since stiffer symmetry energy generally
causes more neutrons to emit in the direction
perpendicular to the reaction plane, larger values of
elliptic flow ratios of neutron and proton
$V_{2}^{n}/V_{2}^{p}$ are seen with stiffer symmetry energies.
With the SRC of nucleon-nucleon, values of the $V_{2}^{n}/V_{2}^{p}$ ratio
are larger than that without the SRC. Fig.~\ref{v2rnp}
indicates the FOPI-LAND elliptic flow experimental data does not favor very
soft symmetry energy ($x$ = 2, 3). Combining the studies of nucleon elliptic flow and  $\pi^{-}/\pi^{+}$ ratio, one can roughly obtain the symmetry energy stiffness parameter $x$ = 1, i.e., a mildly soft density-dependent symmetry energy at supra-saturation densities is obtained as shown in Fig.~\ref{esym}.

\section{Probing the curvature of nuclear symmetry energy $K_{\rm{sym}}$ around saturation density}

The density-dependent symmetry energy at saturation can be Taylor expanded as \cite{issac2009},
\begin{equation}\label{esym2}
E_{sym}(\rho)=E_{sym}(\rho_0)+L\left(\frac{\rho-\rho_0}{3\rho_0}\right)
+\frac{K_{sym}}{2}\left(\frac{\rho-\rho_0}{3\rho_0}\right)^2,
\end{equation}
where $E_{sym}(\rho_0)$ is the value of the symmetry energy at saturation and the quantities $L$, $K_{sym}$ are, respectively, its slope and curvature at saturation,
\begin{equation}
\begin{array}{c}
\displaystyle{L=3\rho_0\frac{\partial E_{sym}(\rho)}{\partial \rho} \Big |_{\rho=\rho_0}} \ , \,\,\,\,
\displaystyle{K_{sym}=9\rho_0^2\frac{\partial^2E_{sym}(\rho)}{\partial \rho^2} \Big |_{\rho=\rho_0}}. \end{array}
\end{equation}
The most probable magnitude $E_{\rm sym}(\rho_0)= 31.7\pm 3.2$ MeV and slope $L= 58.7\pm 28.1 $ MeV of the nuclear symmetry energy at saturation have been obtained \cite{Oertel17,lihan13}. While the curvature $K_{\rm{sym}}$ is still unknown, probably to be in the range of $-400 \leq K_{\rm{sym}} \leq 100$ MeV \cite{Tews17,Zhang17}.
\begin{figure}[tbh]
\centering
\includegraphics[width=0.47\textwidth]{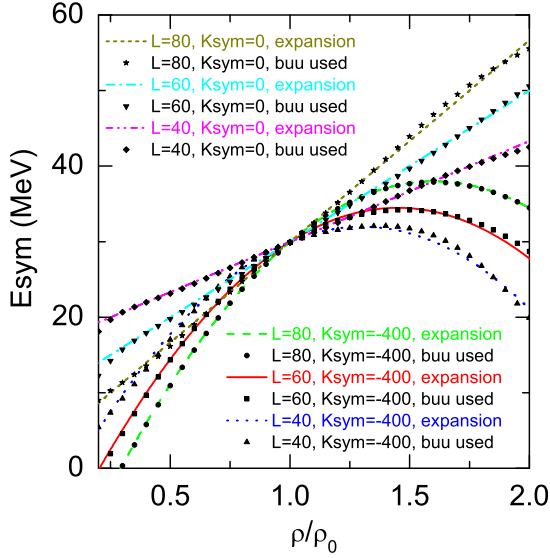}
\caption{Density-dependent symmetry energies for six combinations of curvatures $K_{sym}= -400, 0$ MeV and slopes $L= 40, 60, 80$ MeV obtained from single nucleon potential Eq.~(\ref{buupotential}) (symbols) and Taylor expansion Eq.~(\ref{esym2}) (lines).} \label{esymlk19}
\end{figure}
Fig.~\ref{esymlk19} shows density-dependent symmetry energies derived from single nucleon potential (Eq.~(\ref{buupotential})) in the transport model and that of Taylor expansion (Eq.~(\ref{esym2})). It is seen that the symmetry energies obtained from different combinations of slope values with $L= 40, 60, 80$ MeV and curvatures $K_{sym}= -400, 0$ MeV cover current uncertain range of the high-density symmetry energy as shown in Fig.~\ref{unesym}. The curvature of nuclear symmetry energy not only connects with the high-density symmetry energy, but also closely related to the incompressibility of neutron-rich matter, while it still remains an open problem \cite{issac2009,pie2009}.

\begin{figure}[tbh]
\centering
\includegraphics[width=0.5\textwidth]{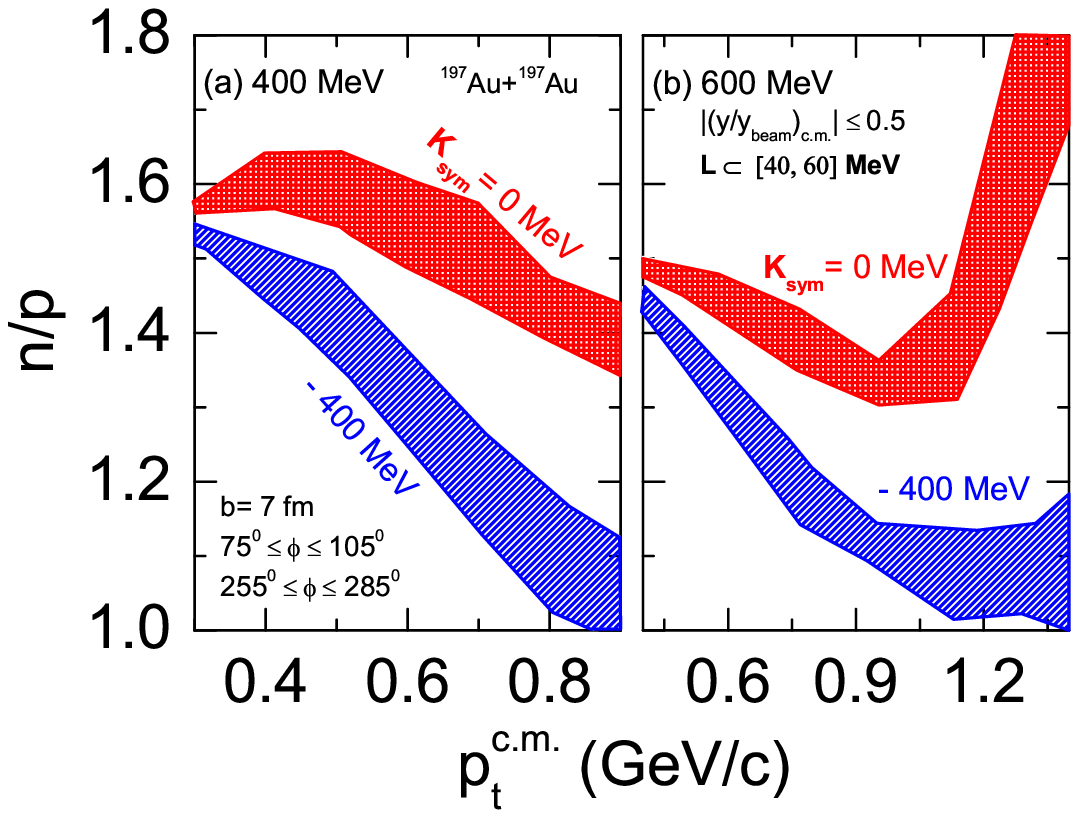}
\caption{Effects of the curvature of nuclear symmetry energy on the squeezed-out neutron to proton ratio in the semi-central reaction of Au+Au at 400 (panel (a)) and 600 (panel (b))
MeV/nucleon. The bandwidths denote uncertainties of the slope $L$. Taken from Ref. \cite{guoyong19}.} \label{squeezernp}
\end{figure}
\begin{figure}[tbh]
\centering
\includegraphics[width=0.4\textwidth]{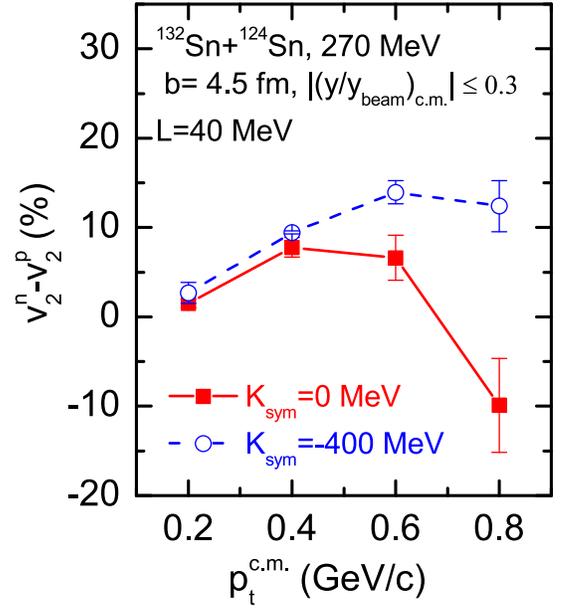}
\caption{Effects of the curvature of nuclear symmetry energy on the difference of neutron and proton elliptic flow in the semi-central reaction $^{132}\rm {Sn}+^{124}\rm {sn}$ reaction at 270 MeV/nucleon.} \label{v2ksym}
\end{figure}
Squeezed-out nucleon, which emitted in the direction perpendicular to the reaction plane in semi-central collisions, is known to carry direct information about the high density phase  \cite{greiner,bert88,cas90,aich91,Reisdorf97,pawl2002,yong07}. Fig.~\ref{squeezernp} shows the squeezed-out neutron to proton ratio in Au+Au at 400 (panel (a)) and 600 (panel (b)) MeV/nucleon \cite{guoyong19}. It is demonstrated that the effects of the curvature of the symmetry energy on the squeezed-out neutron to proton ratio n/p are quite evident, especially at high transverse momenta. The values of the squeezed-out n/p with $K_{sym}= 0$ are higher than that with $K_{sym}= -400$ MeV. It is also seen that, the effects of the curvature of the symmetry energy on the squeezed-out n/p are much larger than the ratio of integrating neutron and proton elliptic flows \cite{cozma}. Fig.~\ref{v2ksym} shows the effects of the curvature of nuclear symmetry energy on the difference of neutron and proton elliptic flow in the semi-central reaction of $^{132}\rm {Sn}+^{124}\rm {sn}$ at 270 MeV/nucleon. It is clearly seen that the difference of neutron and proton elliptic flow is also quite sensitive to the curvature of nuclear symmetry energy.
The isospin-dependent squeezed-out nucleon emission thus may be one potential probe of the high-density symmetry energy.

\section{Summary and perspective}

Nuclear symmetry energy is an old but fundamental and critical physical quantity in isospin nuclear physics, especially in astrophysics relating to neutron stars's production, evolution and merger \emph{etc}. Due to small asymmetry of compressed nuclear matter formed in general heavy-ion collisions in terrestrial laboratory, the effects of nuclear symmetry energy on many observables are generally less than 10 or 20\%. Therefore uncertainties from transport models or some other unclear physical inputs inevitably affect interpretation of nuclear symmetry energy from experimental data. Topics on model dependence, qualitative observables, sensitive probes, many-observable cross-constraints, probed density-region of sensitive observables and the effects of short-range correlations should be demonstrated before constraining the density-dependent symmetry energy from experimental data comparisons.

Besides constraining the nuclear symmetry energy from heavy-ion collisions \cite{yongm2016,fopi16}, the symmetry energy could be also determined via the studies of neutron star merger \cite{naibo,tsang19sym}. It is fascinating to see the match of constraints of the high-density symmetry energy from heaven \cite{ligo17} and earth \cite{plan1,plan2} in the near future.

\section{Acknowledgments}

We thank Lie-Wen Chen, Xiao-Hua Fan, Yuan Gao, Wen-Mei Guo, Bao-An Li, Qing-Feng Li,
Gao-Feng Wei, Yong-Jia Wang, Zu-Xing Yang, Fang Zhang,
Hong-Fei Zhang and Wei Zuo for collaborations. Also the National Natural Science Foundations of China under Grants No. 11775275, 11375239 and No. 11435014 are greatly acknowledged.

\end{document}